\documentclass[final,5p,times]{elsarticle} 

\usepackage{amssymb}

\usepackage{lineno}

\usepackage{graphicx} 

\journal{Nuclear Inst. and Methods in Physics Research, A}

\begin{document}

\begin{frontmatter}



\title{Radiation Damage Effects on Double-SOI Pixel Sensors for X-ray Astronomy}


\author[tus]{Kouichi~Hagino}
\author[tus]{Keigo~Yarita}
\author[tus]{Kousuke~Negishi}
\author[tus]{Kenji~Oono}
\author[tus]{Mitsuki~Hayashida}
\author[tus]{Masatoshi~Kitajima}
\author[tus]{Takayoshi~Kohmura}
\author[kyoto]{Takeshi~G.~Tsuru}
\author[kyoto]{Takaaki~Tanaka}
\author[kyoto]{Hiroyuki~Uchida}
\author[kyoto]{Kazuho~Kayama}
\author[kyoto]{Yuki~Amano}
\author[kyoto]{Ryota~Kodama}
\author[miyazaki]{Ayaki~Takeda}
\author[miyazaki]{Koji~Mori}
\author[miyazaki]{Yusuke~Nishioka}
\author[miyazaki]{Masataka~Yukumoto}
\author[miyazaki]{Takahiro~Hida}
\author[kek]{Yasuo~Arai}
\author[kek2]{Ikuo~Kurachi}
\author[qst]{Tsuyoshi~Hamano}
\author[qst]{Hisashi~Kitamura}

\address[tus]{Department of Physics, School of Science and Technology, Tokyo University of Science, 2641 Yamazaki, Noda, Chiba 278-8510, Japan}
\address[kyoto]{Department of Physics, Faculty of Science, Kyoto University, Kitashirakawa-Oiwakecho, Sakyo-ku, Kyoto 606-8502, Japan}
\address[miyazaki]{Department of Applied Physics, Faculty of Engineering, University of Miyazaki, 1-1 Gakuen-Kibanodai-Nishi, Miyazaki, Miyazaki 889-2192, Japan}
\address[kek]{Institute of Particle and Nuclear Studies, High Energy Accelerator Research Organization (KEK), 1-1 Oho, Tskuba, Ibaraki 305-0801, Japan}
\address[kek2]{Department of Advanced Accelerator Technologies, High Energy Accelerator Research Organization (KEK), 1-1 Oho, Tskuba, Ibaraki 305-0801, Japan}
\address[qst]{National Institute of Radiological Sciences, National Institutes for Quantum and Radiological Science and Technology, 4-9-1 Anagawa, Inage-ku, Chiba-City, Chiba, 263-8555, Japan}

\begin{abstract}
The X-ray SOI pixel sensor onboard the FORCE satellite will be placed in the low earth orbit and will consequently suffer from the radiation effects mainly caused by geomagnetically trapped cosmic-ray protons. Based on previous studies on the effects of radiation on SOI pixel sensors, the positive charges trapped in the  oxide layer significantly affect the performance of the sensor. To improve the radiation hardness of the SOI pixel sensors, we introduced a double-SOI (D-SOI) structure containing an additional middle Si layer in the oxide layer. The negative potential applied on the middle Si layer compensates for the radiation effects, due to the trapped positive charges. Although the radiation hardness of the D-SOI pixel sensors for applications in high-energy accelerators has been evaluated, radiation effects for astronomical application in the D-SOI sensors has not been evaluated thus far. To evaluate the radiation effects of the D-SOI sensor, we perform an irradiation experiment using a 6-MeV proton beam with a total dose of $\sim 5{\rm ~krad}$, corresponding to a few tens of years of in-orbit operation. This experiment indicates an improvement in the radiation hardness of the X-ray D-SOI devices. On using an irradiation of 5~krad on the D-SOI device, the energy resolution in the full-width half maximum for the 5.9-keV X-ray increases by $7\pm2\%$, and the chip output gain decreases by $0.35\pm0.09\%$. The physical mechanism of the gain degradation is also investigated; it is found that the gain degradation is caused by an increase in the parasitic capacitance due to the enlarged buried n-well.
\end{abstract}




\end{frontmatter}


\section{Introduction}
\label{sec:intro}
We propose a future wide-band X-ray astronomical satellite FORCE (Focusing On the Relativistic universe and Cosmic Evolution)~\cite{Mori2016,Nakazawa2018}. The FORCE mission aims to trace cosmic formation history by observing high energy phenomena in the universe.
For this purpose, it carries three pairs of an X-ray super-mirror and a focal-plane detector with a focal length of 10~m. The mirror is composed of thin Si substrates with multi-layer coatings, thereby achieving a high angular resolution in hard X-ray with low mass~\cite{Zhang2018}. The focal-plane detector, which is termed as wideband hybrid X-ray imager (WHXI), comprises a stack of Si sensors and CdTe sensors, utilizing the same concept as the hard X-ray imager onboard Hitomi~\cite{Sato2016,Nakazawa2018a,Hagino2018}. A combination of the light-weight Si mirror and the hybrid detector provides a wide energy coverage from 1 to 80~keV and a high angular resolution of $<15''$.

We have been developing X-ray pixel sensors termed as ``XRPIX'' for the Si sensor of WHXI~\cite{Tsuru2018}. XRPIX is a monolithic active pixel sensor composed of a Si sensor layer and a CMOS pixel circuit layer with a thin oxide layer (BOX: buried oxide) in between. The sensors were fabricated using silicon-on-insulator (SOI) technology, which enables using high and low resistivity Si wafers as the sensor and circuit layers, respectively. Owing to the high resistivity Si wafer, XRPIX has a depletion layer with a thickness of a few hundreds of ${\rm \mu m}$. In the CMOS circuit layer, each pixel circuit has a self-trigger function. This enables a timing resolution of $\sim10{\rm ~\mu s}$.

Radiation hardness is one of the major challenges when developing SOI pixel sensors. SOI pixel sensors are sensitive to the total ionization dose (TID) effect~\cite{Hara2019}. The TID effect is caused by the accumulated positive charges in the BOX layer. Under ionizing irradiation, electron-hole pairs are created in this BOX layer. As a fraction of these holes are trapped in the BOX, it forms a positive oxide-trap charge~\cite{Schwank2008}. These accumulated charges affect the CMOS circuit layer and alter transistor characteristics such as threshold voltage and trans-conductance~\cite{Hara2019}.

The double SOI (D-SOI) structure was introduced to reduce the TID effect. The D-SOI device has an additional thin middle Si layer in the BOX layer~\cite{Miyoshi2013}. This middle Si layer compensates for the positive potential induced by the accumulated charges by applying a small negative voltage ($-2.5{\rm ~V}$ in this work). In addition to TID compensation, the D-SOI structure is useful in terms of spectral performance. In the single SOI (S-SOI) structure, which is the conventional SOI structure, there is a capacitive coupling between the sense node in the sensor layer and the circuit layer. As the middle Si layer is biased at a fixed bias voltage, it acts as an electrostatic shield, and reduces this capacitive coupling~\cite{Ohmura2016}. As a consequence, the chip output gain increased and the readout noise reduced in the D-SOI device~\cite{Takeda2019}.

Although the radiation hardness for D-SOI devices applied in high energy accelerators has been evaluated~\cite{Honda2014,Hara2015,Hara2019}, the devices used for astronomical applications have not been evaluated thus far. Therefore, we irradiated a 6~MeV proton beam on the D-SOI type XRPIX and evaluated its radiation hardness, from a perspective of astronomical applications. The proton irradiation experiment is described in Section~\ref{sec:exp} and its results are reported in Section~\ref{sec:result}. In Section~\ref{sec:discussion}, we discuss the physical mechanism of gain degradation, and the conclusions of this study are summarized in Section~\ref{sec:conclusion}.

\section{Proton Irradiation Experiment}
\label{sec:exp}

\begin{figure}[tbp]
\centering
\includegraphics[width=\hsize]{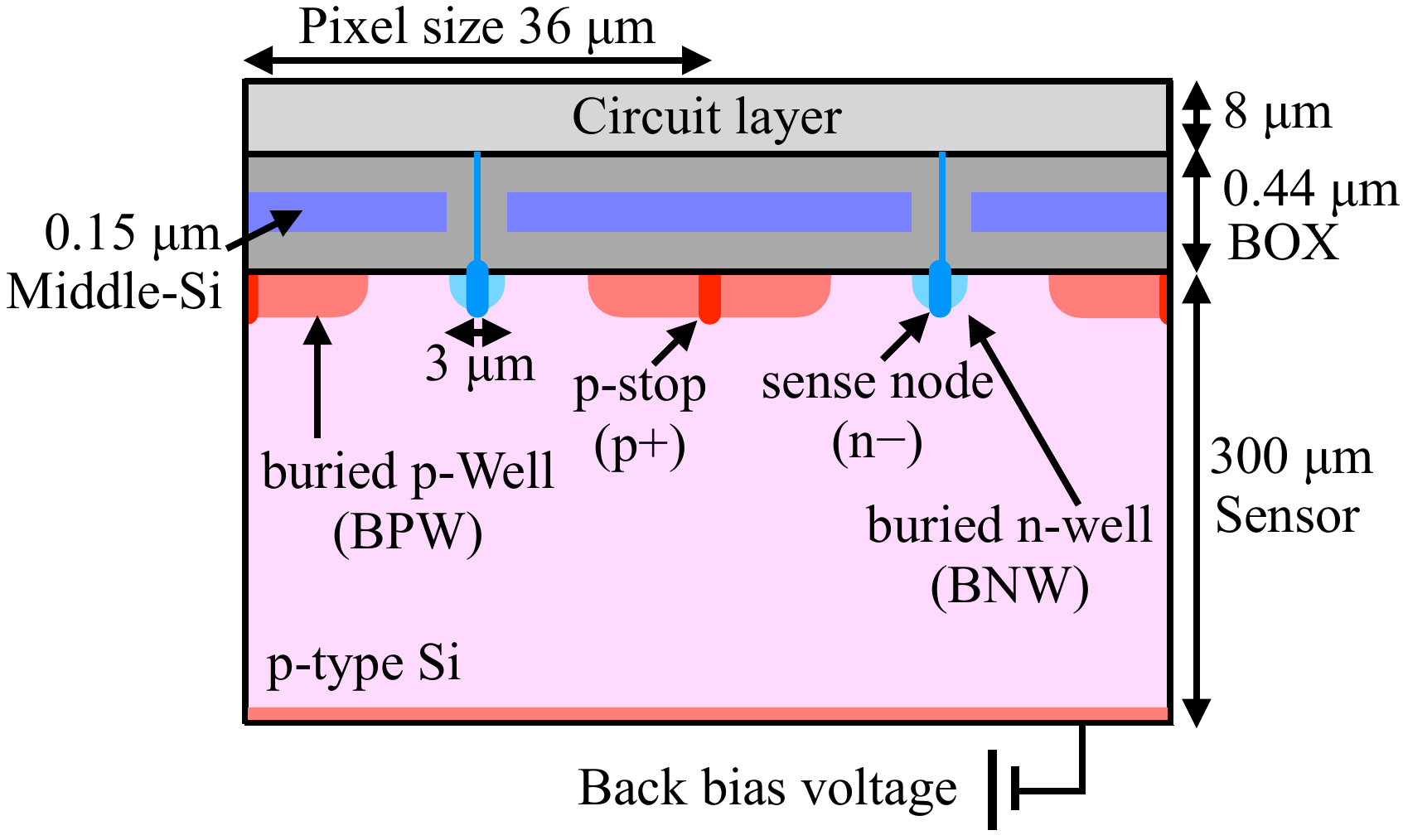}
\caption{Schematic picture of XRPIX6C}
\label{fig:xrpix6c}
\end{figure}

\subsection{Radiation Hardness Required for FORCE}
In the orbit of the FORCE satellite (altitude of $\sim500{\rm~km}$ and orbital inclination of $\sim30^\circ$), the onboard sensors experience radiation damage mainly due to the cosmic-ray protons geomagnetically trapped at South Atlantic Anomaly. The dose rate of the trapped protons on the XRPIX is approximately $\sim0.1{\rm ~krad/year}$~\cite{Yarita2018}. As X-ray astronomical satellites operate for a few to $\sim10$~years in orbit, the total dose during the mission lifetime is less than a few krad.

Compared with high energy accelerators, the typical dose level in astronomical satellites is significantly lower. However, in terms of device performance, more stringent requirements are imposed in astronomical satellites. XRPIX is required to have an energy resolution of $300{\rm ~eV}$ at $6{\rm ~keV}$ in full-width half maximum (FWHM) and a readout noise of $10{\rm ~e^{-}}$ in root mean square (rms)~\cite{Tsuru2018}. Thus, we investigate the degradation of spectral and noise performances with an accuracy of $\sim10{\rm ~eV}$ (a few ${\rm e^{-}}$) under a radiation dose of a few krad.

\subsection{D-SOI device: XRPIX6C}
We performed a proton irradiation experiment on the D-SOI device called ``XRPIX6C,'' at Heavy Ion Medical Accelerator in Chiba (HIMAC), in the National Institute of Radiological Sciences. A schematic of the cross-sectional structure of XRPIX6C is depicted in Fig.~\ref{fig:xrpix6c}. The sensor layer is composed of p-type Si with a resistivity of $4{\rm ~k\Omega~cm}$, which corresponds to the doping concentration of $3\times 10^{12}{\rm ~cm^{-3}}$. As the thickness is $300{\rm ~\mu m}$, the back bias voltage should be higher than $\sim200{\rm ~V}$ for full depletion. The pixel size is $36{\rm ~\mu m}\times 36{\rm ~\mu m}$. The size of imaging area is $1.728\times1.728{\rm ~mm^2}$ in this device, and will be $15\times 45{\rm ~mm^2}$ in the flight model~\cite{Tsuru2018}. The power consumption of this device is $\sim 1{\rm ~W}$, including the readout board. Each pixel has a sense node surrounded by a buried n-well (BNW), which was originally introduced as an electric shield~\cite{Arai2011}. In the D-SOI device, as the middle Si layer acts as an electric shield, the size of the BNW is as small as $3{\rm ~\mu m}$, which is significantly smaller than that of the BNW in S-SOI. Owing to the smaller BNW size in the D-SOI device, the parasitic capacitance of the sense node is significantly reduced, and the signal-to-noise ratio is improved~\cite{Takeda2019}.

XRPIX6C was operated during proton irradiation because the device will be operated in the space radiation environment. The sensor layer was fully depleted by applying a back bias voltage of $-250{\rm ~V}$, and the readout circuits were operated as usual. A negative voltage of $-2.5{\rm ~V}$ was applied to the middle Si layer. The device was placed in a vacuum chamber and cooled down to $\sim -70^\circ{\rm C}$, which is determined by the experimental setup of the cooler and the chamber. Although this temperature is lower than that in the space environment ($\sim -15^\circ{\rm C}$), the noise due to the leakage current is negligible even at $-15^\circ{\rm C}$.

\subsection{Experimental Setup}

\begin{figure}[tbp]
\centering
\includegraphics[width=\hsize]{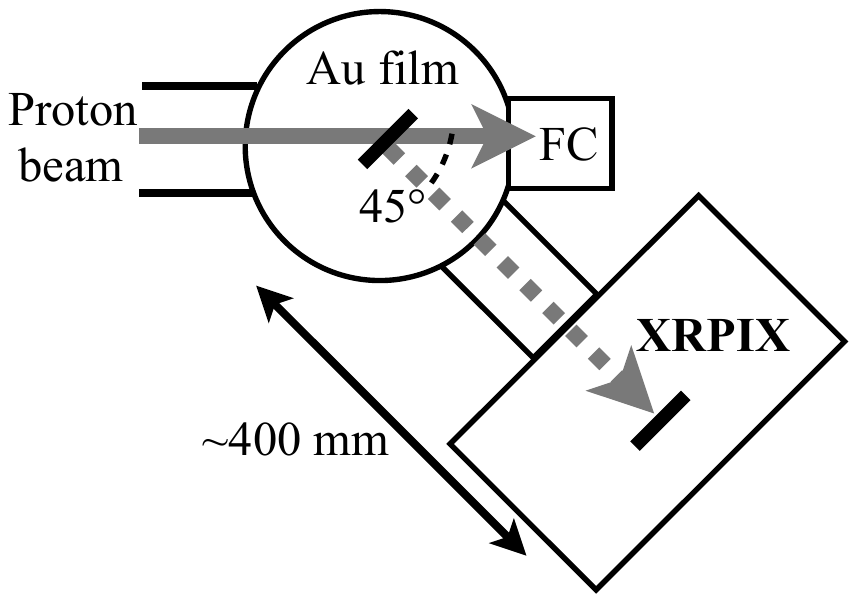}
\caption{Schematic of the proton irradiation experiment. XRPIX is irradiated with a proton beam scattered to $45^\circ$.}
\label{fig:setup}
\end{figure}

\begin{figure*}[tbp]
\centering
\includegraphics[width=0.48\hsize]{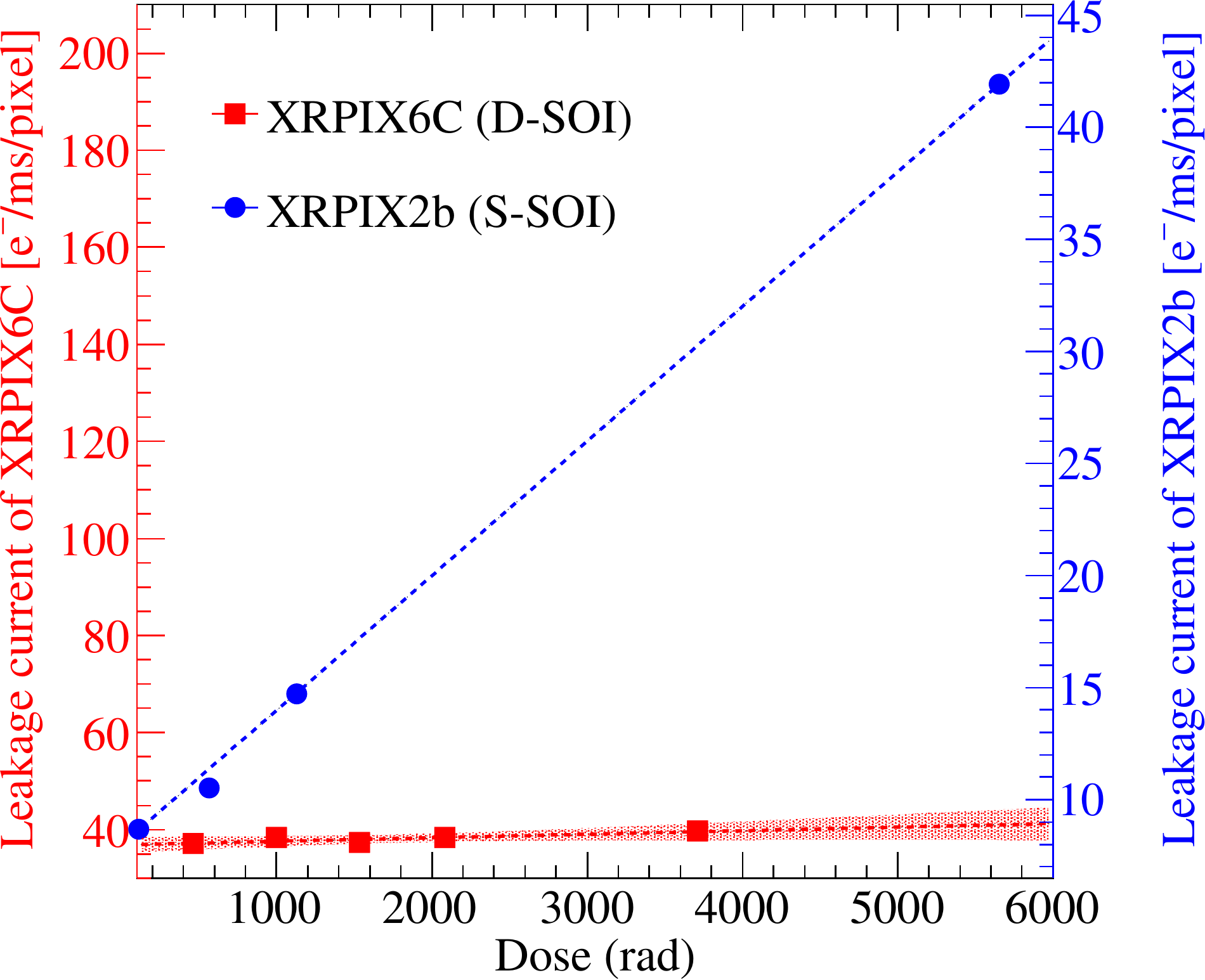}
\hspace{0.015\hsize}
\includegraphics[width=0.48\hsize]{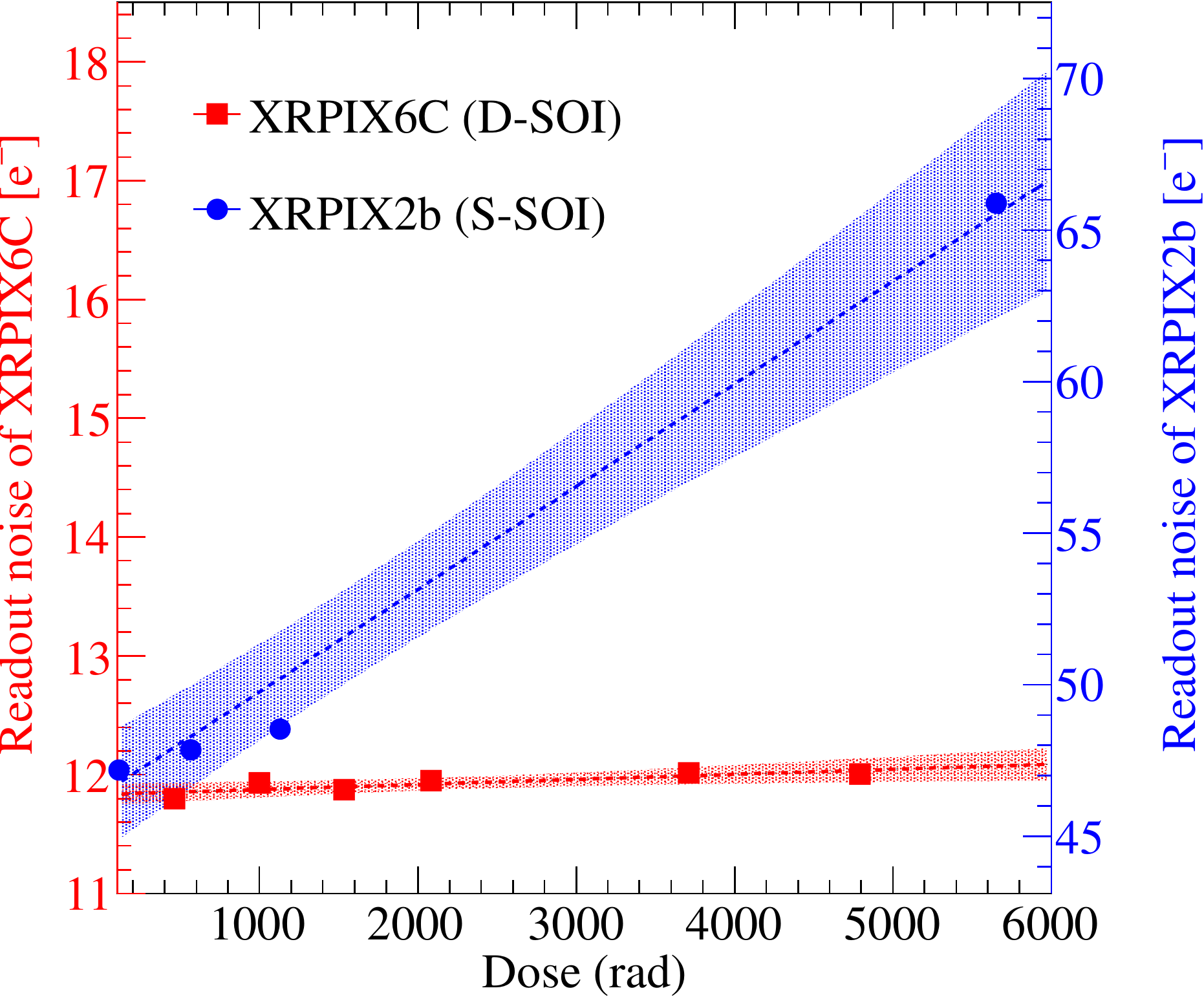}
\caption{Degradation of leakage current (left panel) and readout noise (right panel) of XRPIX6C (D-SOI device; red color/left axis) compared with those of XRPIX2b (S-SOI device; blue color/right axis). Vertical axes are scaled to match each other at non-irradiation. The lines and shades indicate the best fit linear function and its 90\% confidence region, respectively. All the plots for XRPIX2b are the re-analysis results of the experimental data in \citet{Yarita2018}.}
\label{fig:leak_noise}
\end{figure*}

The experimental setup is shown in Fig.~\ref{fig:setup}. The proton beam is scattered by a thin ($2.5{\rm ~\mu m}$)-gold film. XRPIX is installed at a scattered angle of $45^\circ$ at a distance of $\sim400{\rm ~mm}$ from the scatterer. One of the main advantages of this configuration is the spatial uniformity of the beam at the location of XRPIX. The device size of $\sim 1{\rm ~mm}$ corresponds to the angle difference $\Delta \theta=0.1^\circ\textrm{--}0.2^\circ$ from the scatterer. As the differential cross section of the Rutherford scattering is written as $d\sigma/d\Omega\propto \sin^{-4}\left(\theta/2\right)$, the non-uniformity of the beam flux due to the angle difference will be a few percent. In addition, by measuring the unscattered beam using a Faraday cup, the fluctuations in the beam intensity can be monitored during the irradiation.

The energy of the incident proton beam is $6{\rm ~MeV}$ at the beam line. Although the spectrum of the geomagnetically-trapped protons in the FORCE orbit is a continuous spectrum peaked at $\sim 100{\rm ~MeV}$, the energy deposit on the Si sensor is dominated by $4\textrm{--}20{\rm ~MeV}$ protons~\cite{Yarita2018}. Therefore, $6{\rm ~MeV}$ is a good approximation of the radiation environment in the orbit. In addition, since this energy is sufficient to penetrate the BOX layer of XRPIX6C, the vertical non-uniformity of the dose level in the BOX layer is negligible. The beam intensity at the location of XRPIX was measured using an avalanche photodiode; it was found to be $0.84\textrm{--}1.52\times 10^5{\rm ~protons/s/cm^2}$, corresponding to a dose rate of $0.26\textrm{--}0.47{\rm ~krad/hour}$, assuming a stopping power of SiO$_2$ of $53.85{\rm ~MeV~cm^2/g}$ from the NIST PSTAR database~\cite{PSTAR}. The range of values denotes day-by-day variation of the beam intensity. Since such a variation was measured with the Faraday cup during the irradiation, it was properly corrected in the total dose estimation. 
XRPIX6C was intermittently irradiated with the proton beam up to a total dose level of $\simeq 5{\rm ~krad}$, and the device performance between the irradiations was evaluated.

\section{Performance Degradation of XRPIX6C}
\label{sec:result}
\subsection{Leakage Current and Readout Noise}
\begin{figure}[tb]
\centering
\includegraphics[width=\hsize]{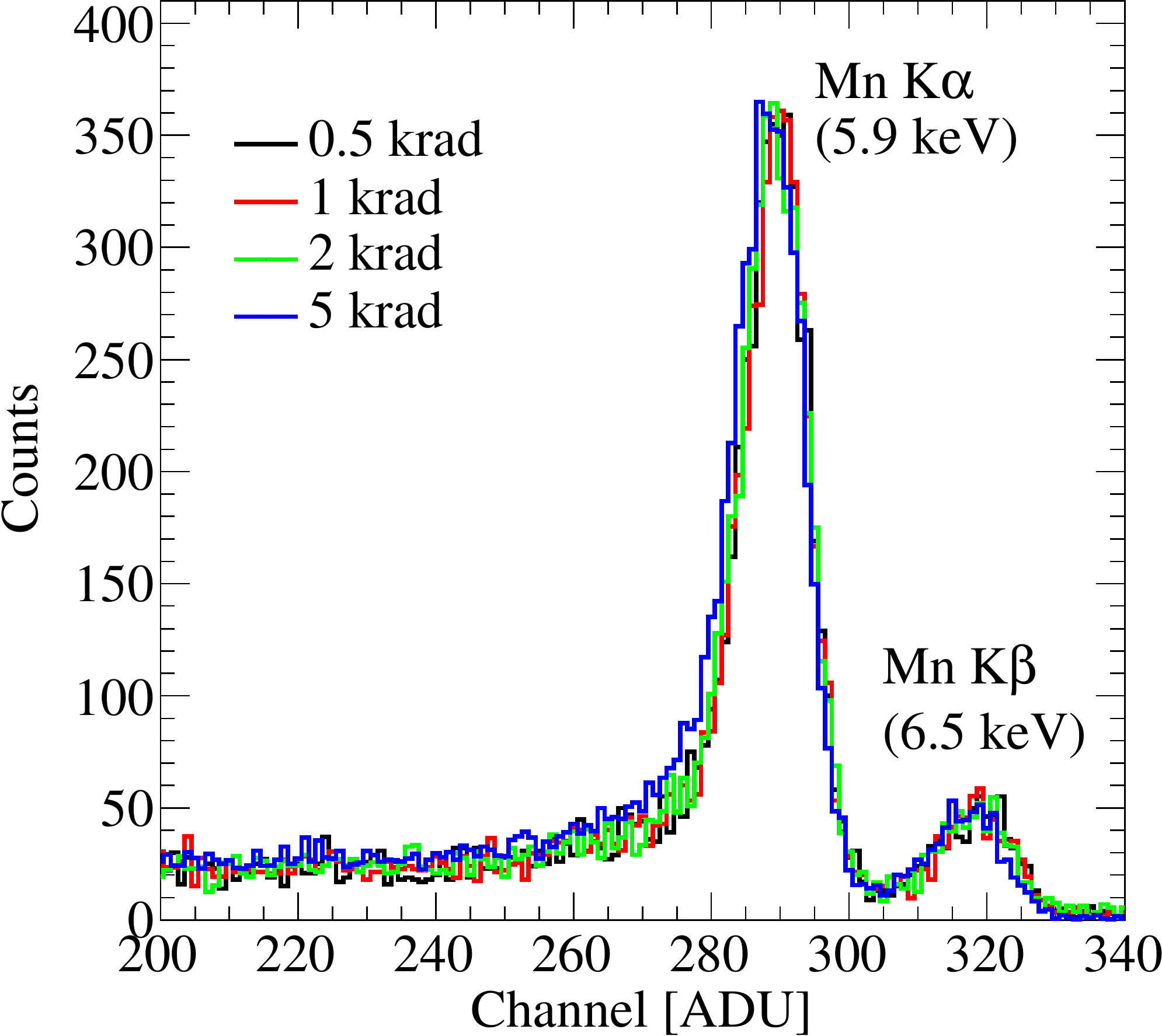}
\caption{X-ray spectra of ${\rm Mn~K\alpha}$ and ${\rm Mn~K\beta}$ from $^{55}{\rm Fe}$ measured using the D-SOI device after proton irradiation.}
\label{fig:spec}
\end{figure}

\begin{figure*}[tb]
\centering
\includegraphics[width=0.48\hsize]{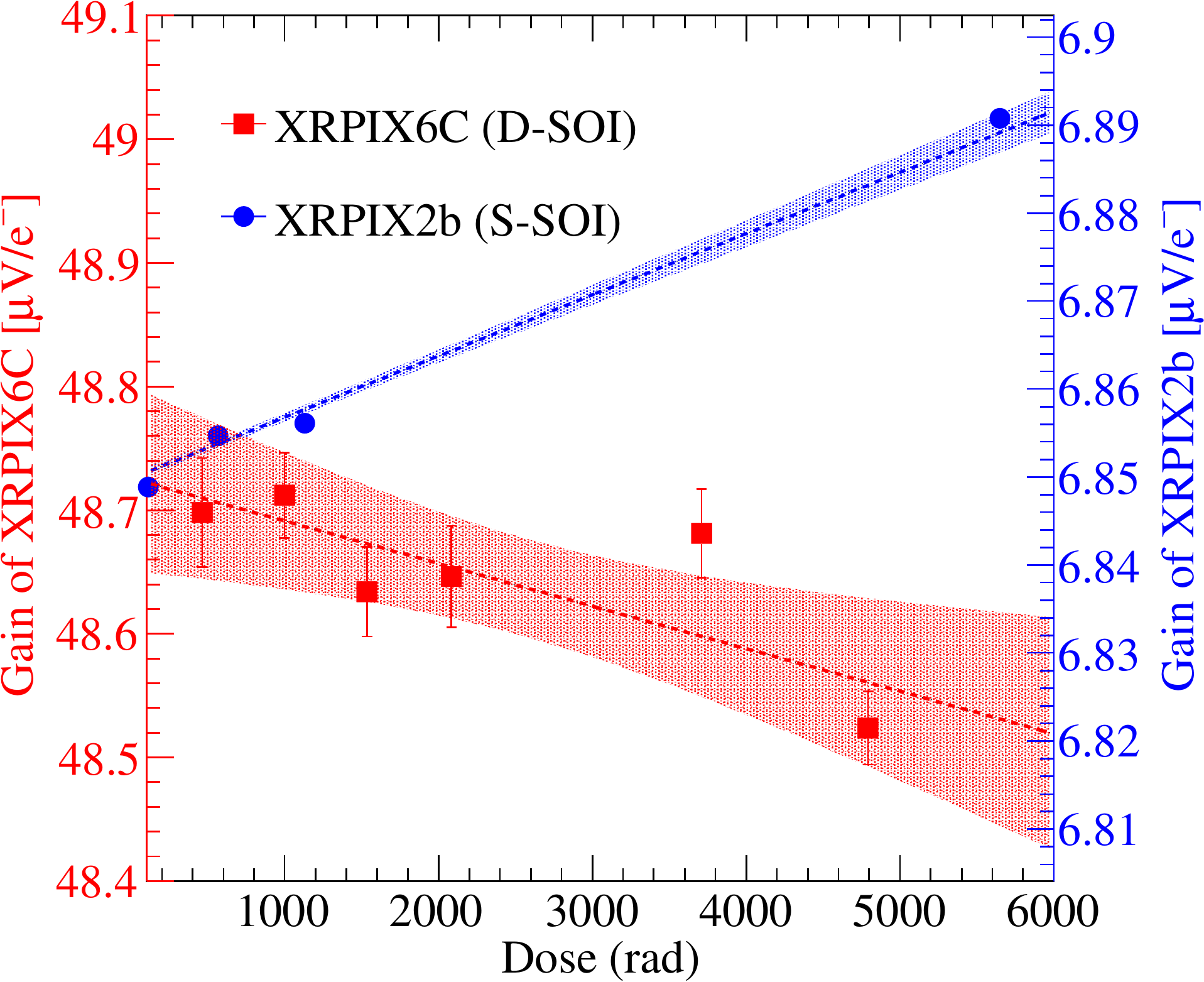}
\hspace{0.015\hsize}
\includegraphics[width=0.48\hsize]{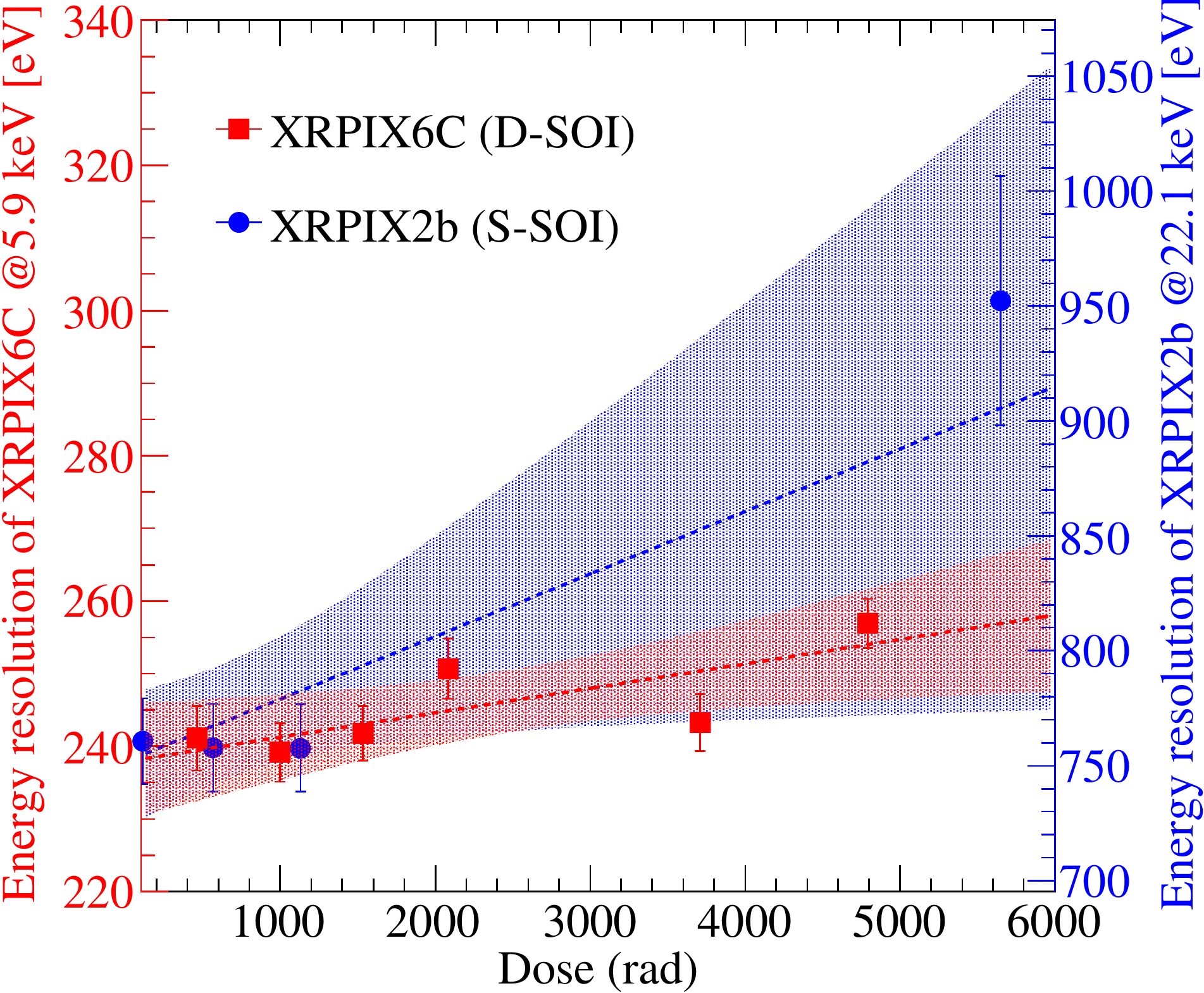}
\caption{Degradation of chip output gain (left panel) and energy resolution (right panel) of XRPIX6C (D-SOI device), compared with those of XRPIX2b (S-SOI). The results of S-SOI and the fitted functions are overplotted in the same manner as in Fig.~\ref{fig:leak_noise}.}
\label{fig:gain_fwhm}
\end{figure*}

We evaluated the degradation in the leakage current by analyzing the pedestal values. In the pixel circuits of XRPIX, charges collected at the sense node are stored in sampling capacitors during a predefined integration time. As the leakage current also flows into the sampling capacitor, the pedestal values should be proportional to the leakage current accumulated during the integration time. Thus, we measured the pedestal values as a function of the integration time and subsequently estimated the leakage current based on its derivative. The left panel of Fig.~\ref{fig:leak_noise} illustrates the degradation in the leakage current. Compared with the result of the S-SOI device, which is indicated in blue~\cite{Yarita2018}, the increase ratio of the leakage current of the D-SOI device is significantly lesser. Depending on the linear fitting, the leakage current of the D-SOI device increases by $9.9\pm4.0\%$ for an irradiation of $5{\rm ~krad}$. We should note that the reason of the large initial leakage current of the D-SOI device is not resolved.

The readout noise was also evaluated based on the pedestal values. We fitted the histogram of the pedestal values for all pixels using a Gaussian function, and its standard deviation was used as a measure of the readout noise. The right panel of Fig.~\ref{fig:leak_noise} depicts the degradation in the readout noise. Similar to the leakage current, the degradation in the readout noise of the D-SOI device is significantly reduced. The best fit line indicates that the readout noise increases by $1.8\pm 0.5\%$ for an irradiation of $5{\rm ~krad}$ in the D-SOI device. Moreover, the readout noise is improved owing to the electrical shielding due to the middle Si layer as described in Sec.~\ref{sec:intro}; it remains comparable with the requirement of $\sim 10{\rm ~eV}$ even after irradiation.

\subsection{X-ray Spectral Performance}
The spectral performance was evaluated by irradiating X-rays from the $^{55}$Fe radioisotope. As shown in Fig.~\ref{fig:spec}, the spectral performance changes slightly after a few krad of dose. For a more quantitative evaluation, we fitted the ${\rm Mn~K\alpha}$ line at 5.9~keV, using a Gaussian function. The chip output gain and the energy resolution were estimated based on the peak position and FWHM of the best-fit Gaussian function, respectively. In this paper, the chip output gain is described in the unit of $\rm \mu V/e^{-}$, meaning the conversion coefficient of the signals from the Si sensor ($\rm e^{-}$) to charge signals in the chip output ($\rm \mu V$).

Degradations in the chip output gain and energy resolution are depicted in Fig.~\ref{fig:gain_fwhm}. Unlike the leakage current and readout noise, the gain and energy resolution do not exhibit a significant increase or decrease in the D-SOI device. At $5{\rm ~krad}$, the gain decreases by $0.35\pm0.09\%$, and the energy resolution degrades by $7.1\pm2.2\%$. Compared with the results of S-SOI, the gain of D-SOI degrades in the opposite direction. The gain of the D-SOI device decreases, whereas that of the S-SOI device increases. Although the increase ratio of the energy resolution improves, this increased value is evidently different between the D-SOI and S-SOI devices. The energy resolution of the D-SOI device increases by $\sim 20{\rm ~eV}$, whereas that of the S-SOI device increases by $\sim 200{\rm ~eV}$. This increased value does not change even if we consider the energy dependence of the energy resolution (i.e., Fano noise) because the Fano noises are as small as 120~eV and 230~eV at 5.9~keV (D-SOI XRPIX6C) and 22.1~keV (S-SOI XRPIX2b), respectively. Therefore, even after an irradiation of $5{\rm ~krad}$, the energy resolution is $\simeq260{\rm ~eV}$, which satisfies the requirement of the FORCE satellite.

\section{Discussion on the Degradation Mechanism}
\label{sec:discussion}

\subsection{Device Simulation}
The proton irradiation experiment of the D-SOI device revealed the degradation of spectral performance. In particular, the gain degrades by a few tens of eV for the S-SOI device as well as the D-SOI device. In addition, the gain of these devices degrades in opposite directions, which would be an insightful feature. Thus, in this section, we discuss the degradation mechanism of the chip output gain in the D-SOI device.

To investigate the mechanism of gain degradation, we calculated the electric field structure and carrier distributions in XRPIX6C using the semiconductor device simulator HyDeLEOS, which is a part of the TCAD system HyENEXSS~\cite{TCAD}. The implementation of the device structure was identical to that in \citet{Hagino2019}. The sense nodes, p-stops, BPWs, BNWs, and middle Si layers were implemented based on the parameters provided by LAPIS Semiconductor Co. Ltd. In the simulation, the TID effect was reproduced by setting the positive fixed charges of $\sim 10^{11}{\rm ~cm^{-2}}$ in the BOX layer, based on the previous studies on SOI pixel devices~\cite{Hara2019}.

\begin{figure}[tbp]
\centering
\includegraphics[width=\hsize]{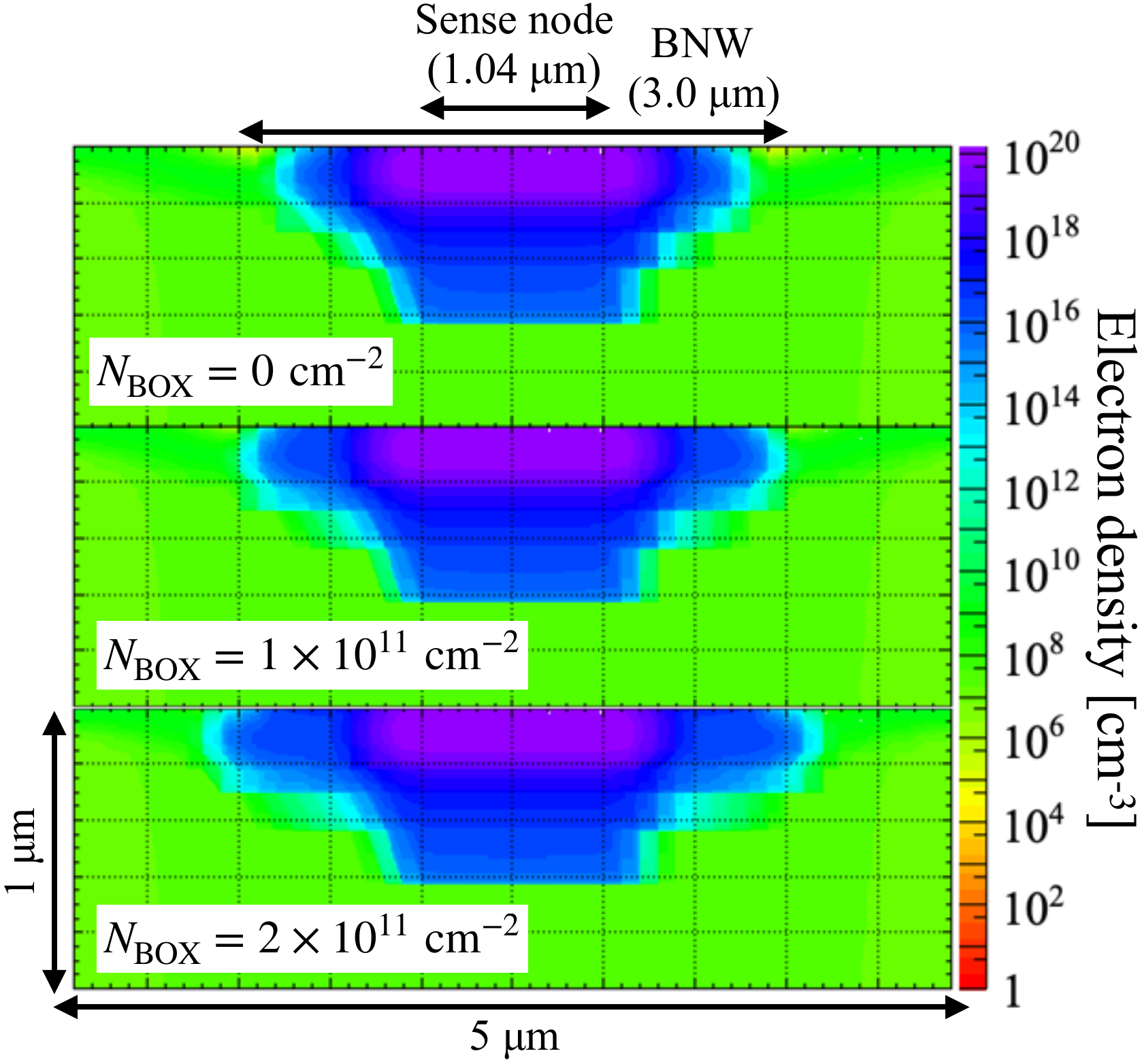}
\caption{Two-dimensional electron density map around the BNW with BOX charges of $0$, $1\times10^{11}$, and $2\times 10^{11}{\rm ~cm^{-2}}$.}
\label{fig:bnw}
\end{figure}

Based on the results of the device simulation, we found that the BNW size varies with the BOX charge $N_{\rm BOX}$. Fig.~\ref{fig:bnw} depicts the electron density map calculated via the TCAD simulation. As shown in the figure, the effective size of the BNW, whose electron concentration is $10^{16\textrm{--}18}{\rm ~cm^{-3}}$, increases from $\simeq2.6{\rm ~\mu m}$ at $N_{\rm BOX}=0{\rm ~cm^{-2}}$ to $\simeq3.4{\rm ~\mu m}$ at $N_{\rm BOX}=2\times10^{11}{\rm ~cm^{-2}}$. This enlargement of the BNW is attributed to the positive potential due to the BOX charges. Such a potential attracts electrons towards the interface between the sensor and the BOX layers, thereby enhancing the electron density at the interface. Thus, the BOX charge generated via irradiation enlarges the BNW. We believe that this effect is one of the causes of gain degradation.

\subsection{Relation of the BNW Size to the Gain:\\ consideration with another D-SOI device XRPIX6D}

\begin{figure}[tbp]
\centering
\includegraphics[width=\hsize]{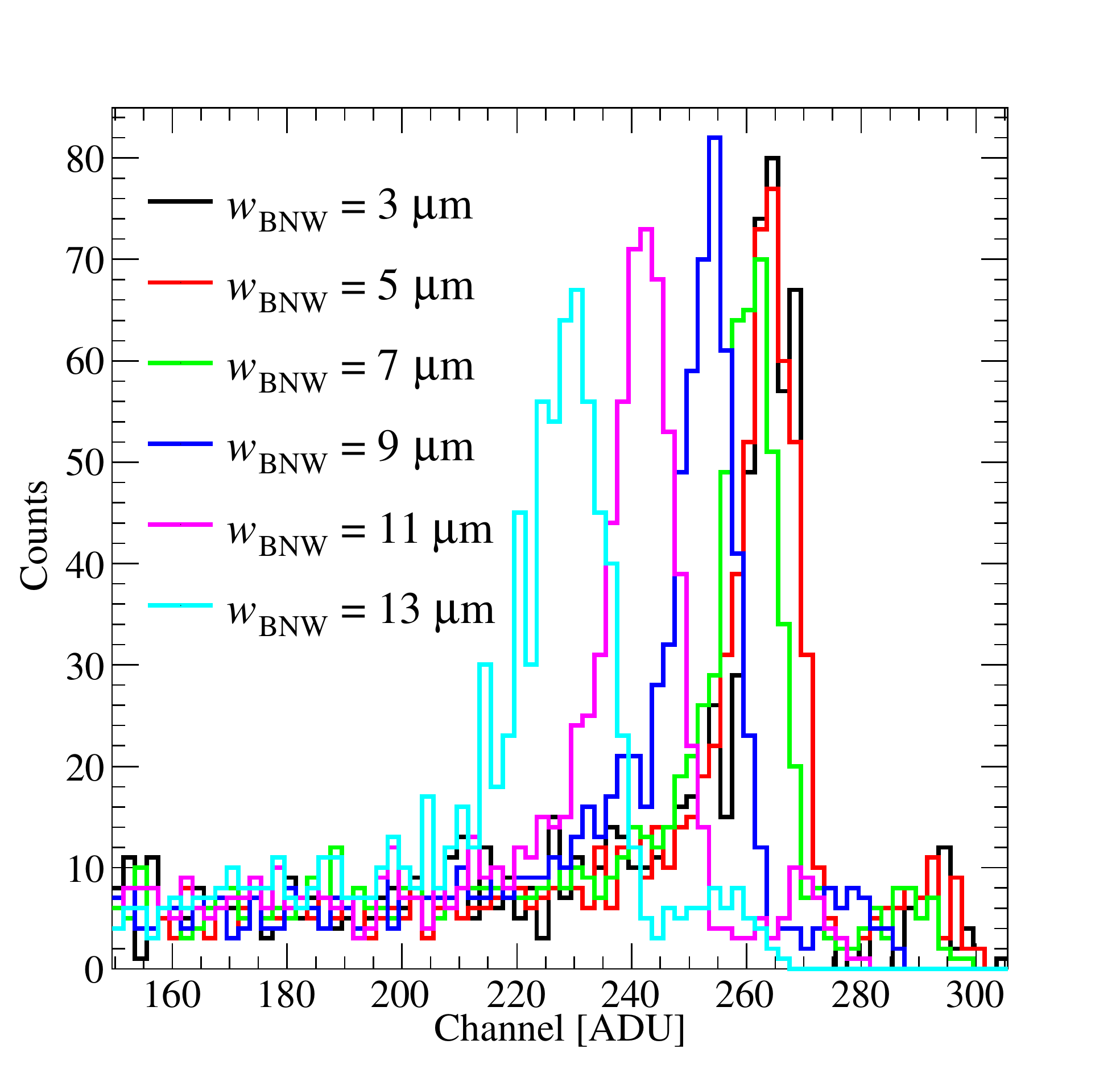}
\caption{X-ray spectra of the D-SOI device with different BNW sizes. The peak position depends on the size of the BNW.}
\label{fig:allteg}
\end{figure}

To investigate the relation between BNW size and chip output gain, we analyzed the experimental data of another D-SOI device called ``XRPIX6D''~\cite{Hagino2019}. Test element groups with different BNW sizes were implemented in this device. The spectra obtained for the different BNW size are depicted in Fig.~\ref{fig:allteg}. The chip output gain decreases with the increasing BNW size. This indicates that BNW enlargement due to the irradiation would probably result in gain degradation.

As a more detailed physical mechanism of gain degradation, we considered the effect of the sense node capacitance on chip output gain. The chip output gain of XRPIX is determined using the closed-loop gain of a charge-sensitive amplifier (CSA) $G_{\rm CSA}$, source follower circuit gain $G_{\rm SF}\simeq0.82$, and gain from sample-hold to the output buffer circuit $G_{\rm SH}\simeq0.8$~\cite{Takeda2019}. The sense node capacitance affects the CSA gain. Considering a CSA circuit with input capacitance $C_{\rm SN}$ (in this case, sense node capacitance) and a feedback capacitance $C_{\rm FB}$, the inverse of chip output gain $G$ can be written as
\begin{equation}
\frac{1}{G}=\frac{1}{G_{\rm CSA}G_{\rm SF}G_{\rm SH}}=\frac{1}{AG_{\rm SF}G_{\rm SH}}C_{\rm SN}+\frac{A+1}{AG_{\rm SF}G_{\rm SH}}C_{\rm FB},
\end{equation}
where $A$ is the open-loop gain of the CSA. If the sense node capacitance increases from $C_{\rm SN}$ to $C_{\rm SN}+\Delta C_{\rm SN}$, the inverse of the gain changes as
\begin{equation}
\Delta \left(\frac{1}{G}\right)=\frac{1}{AG_{\rm SF}G_{\rm SH}}\Delta C_{\rm SN}.\label{eq:dGdC}
\end{equation}
As the parasitic capacitance between the BNW and the middle Si layer is considered to be a major contributor to the sense node capacitance $C_{\rm SN}$, we estimated the parasitic capacitance using the parallel-plate capacitor formula. Based on the distance between the BNW and the middle Si layer $d_{\rm BNW-MS}=0.145{\rm ~\mu m}$, the open-loop gain $A=108$~\cite{Takeda2019}, and a permittivity of SiO$_2$ $\varepsilon=3.5\times 10^{-13}{\rm~F/cm}$, Eq.~\ref{eq:dGdC} is expressed as
\begin{equation}
\Delta \left(\frac{1}{G}\right)\simeq 3.4\times10^{-3}\times \left(\frac{\Delta S_{\rm BNW}}{1{\rm ~\mu m^2}}\right) {\rm ~fF},\label{eq:dGdS}
\end{equation}
where $\Delta S_{\rm BNW}$ is the change in the areas of the BNW.

Utilizing the data of XRPIX6D depicted in Fig.~\ref{fig:allteg}, we verified the validity of Eq.~\ref{eq:dGdS}. In this data, the BNW size changes from $3{\rm ~\mu m}$ to $13{\rm ~\mu m}$, corresponding to $\Delta S_{\rm BNW}\simeq160{\rm ~\mu m^2}$. Thus, the change in the inverse of the gain is calculated to be $\Delta (1/G)\simeq 0.54{\rm ~fF}$, using Eq.~\ref{eq:dGdS}. On the other hand, based on the experimental data of XRPIX6D, the change in the inverse of the gain is calculated to be $\Delta (1/G)\simeq 0.56 {\rm ~fF}$. Thus, the change in the gain corresponding to the change in the size of the BNW can be appropriately explained by the relation in Eq.~\ref{eq:dGdS}.

\subsection{Effect of the BNW Enlargement due to the Irradiation}
Assuming the relation in Eq.~\ref{eq:dGdS}, we calculated the change in the area of the BNW required to explain the gain degradation observed during the proton irradiation experiment. In the experiment, the gain degradation was found to be $0.35\%$ at an irradiation of $5{\rm ~krad}$, corresponding to $\Delta (1/G)\simeq 1.1\times 10^{-2}{\rm ~fF}$. According to Eq.~\ref{eq:dGdS}, to explain this gain degradation, the BNW area should change by $\Delta S_{\rm BNW}\simeq 3.2{\rm ~\mu m^2}$. On the other hand, the simulation of the device indicates that the BNW area changes by $\simeq2\textrm{--}5{\rm ~\mu m^2}$ for the accumulated BOX charge of $1\textrm{--}2\times 10^{11}{\rm ~cm^{-2}}$. According to a previous study~\cite{Hara2019}, this value of the BOX charge is reasonable for an irradiation of $5{\rm ~krad}$. Moreover, the opposite direction of gain degradation in the S-SOI device XRPIX2b can also be explained by this scenario because XRPIX2b consists of an n-type substrate and p-type sense nodes surrounded by BPWs rather than BNW. In the n-type substrate, the electrons attracted to the sensor/BOX interface would shrink the BPW around the sense node, thereby increasing the gain. Thus, the gain degradation observed during the proton irradiation experiment can be quantitatively explained by the increase in the sense node capacitance owing to the enlargement of the BNW.

The gain degradation due to BNW enlargement must also affect the readout noise. In previous studies, the readout noise in XRPIX was found to have a strong correlation with the gain~\cite{Takeda2015,Harada2018}. According to these studies, the readout noise $\sigma$ and the gain $G$ have a power law relation of $\sigma\propto G^{-0.7}$. Thus, the gain degradation of $\Delta G/G\simeq 0.35\%$ observed in our experiment corresponds to the readout noise increase of $\Delta \sigma\simeq 0.3\%$. In addition, the increase of the shot noise due to the degradation of the leakage current also contributes to the increase in the readout noise. As the readout noise was evaluated using an integration time of $1{\rm ~ms}$, the contribution of the increase in the shot noise to the readout noise is estimated to be $\simeq1.3\pm0.5\%$. Thus, the degradation in the readout noise can be completely explained by a combination of the contributions of gain degradation and the increase in the leakage current.

Although the degradation of the gain and readout noise is explained by the above scenario, the degradation mechanism of the energy resolution is not fully understood. One possibility is the charge loss, which was analyzed in detail in our previous study~\cite{Hagino2019}. In the D-SOI devices, a part of the signal charge generated by the incident X-ray is probably lost at the Si/SiO$_2$ interface between the sensor layer and the BOX layer. The charge loss makes a tail structure in the X-ray spectra, and degrades the energy resolution. Since the amount of the charge loss must be affected by the electric field and the carrier distribution in the sensor layer, the BOX charges generated via irradiation could also affect the energy resolution. However, in the current experimental data shown in Fig.~\ref{fig:spec}, it is difficult to evaluate the change of the tail structure. Thus, in order to investigate the radiation effect from this aspect, it is necessary to study at much higher dose level, where the degradation of the energy resolution would be more significant.

\section{Conclusions}
\label{sec:conclusion}
We evaluated the radiation hardness of the new XRPIX with a D-SOI structure, by irradiating a 6-MeV proton beam at HIMAC. We found that the degradation in the leakage current and readout noise was improved in the D-SOI device.
Even after an irradiation of $\sim5{\rm ~krad}$, the energy resolution satisfies the requirement of the FORCE mission ($<300{\rm ~eV}$). Moreover, the gain degradation could be explained by the enlargement in the size of the BNW caused by the BOX charges ($1\textrm{--}2\times10^{11}{\rm ~cm^{-2}}$). The readout noise degradation was also found to be consistent with the effect of gain degradation and the increase in leakage current.

\section*{Acknowledgments}
We acknowledge the valuable advice and assistance provided by the personnel of LAPIS Semiconductor Co., Ltd. This study was supported by MEXT/JSPS KAKENHI Grant-in-Aid for Scientific Research on Innovative Areas 25109002 (Y.A.), 25109004 (T.G.T., T.T., K.M., A.T., and T.K.), Grant-in-Aid for Scientific Research (B) 25287042 (T.K.), Grant-in-Aid for Young Scientists (B) 15K17648 (A.T.), Grant-in-Aid for Challenging Exploratory Research 26610047 (T.G.T.), and Grant-in-Aid for Early-Career Scientists 19K14742 (A.T.). This study was also supported by the VLSI Design and Education Center (VDEC), the University of Tokyo in collaboration with Cadence Design Systems, Inc., Mentor Graphics, Inc., and Synopsys, Inc.




\section*{References}
\bibliographystyle{elsarticle-num-names} 
\bibliography{ref}





\end{document}